\documentclass[10pt]{article}
\usepackage{textcomp}
\usepackage{xcolor}
\usepackage{colortbl}
\usepackage{graphicx}
\usepackage{amsmath}
\usepackage{gensymb}
\usepackage{dcolumn}
\begin{document}

\title{Instability of Near-Extreme Solutions to the Whitham Equation}
\author{John D. Carter}
\date{\today}
\maketitle

\begin{abstract}
    The Whitham equation is a model for the evolution of small-amplitude, unidirectional waves of all wavelengths on shallow water.  It has been shown to accurately model the evolution of waves in laboratory experiments.  We compute $2\pi$-periodic traveling-wave solutions of the Whitham equation and numerically study their stability with a focus on solutions with large steepness.  We show that the Hamiltonian oscillates as a function of wave steepness when the solutions are sufficiently steep.  We show that a superharmonic instability is created at each extremum of the Hamiltonian and that between each extremum the stability spectra undergo similar bifurcations.  Finally, we compare these results with those from the Euler equations.
\end{abstract}

\section{Introduction}

\subsection{The Euler equations}

The irrotational motion of an inviscid, incompressible, homogeneous, two-dimensional fluid with a free surface over a horizontal impermeable bed can be modeled by a system of partial differential equations known as the Euler equations, see for example~\cite{johnson,LannesBook}.  Stokes~\cite{Stokes} established the existence of periodic traveling-wave solutions to the Euler equations and determined the first few terms in series expansions of these solutions using asymptotic Fourier expansions.  These solutions are now referred to as Stokes waves.  Stokes~\cite{StokesSteepestWave} conjectured that the solution with largest wave height on an infinite-depth fluid has a crest angle of $2\pi/3$.  Many years later, Amick {\emph{et al.}}~\cite{EulerSteepestWave} and Plotnikov~\cite{Plotnikov} independently proved this conjecture.  Amick {\emph{et al.}}~\cite{EulerSteepestWave} also proved that the solution with largest height on a finite-depth fluid also has a crest angle of $2\pi/3$. Mitchel~\cite{Mitchel} estimated that the most extreme Stokes wave on a fluid of infinite depth has a steepness of $s=0.142$ where steepness is defined by $s=H/L$ and $H$ and $L$ are the wave height and wavelength of the wave, respectively.  Dyachencko {\emph{et al.}}~\cite{ExtremeStokes} determined that the solution on an infinite-depth fluid with maximal steepness has steepness $s=0.14106348\dots$ using very precise computations.

Benjamin \& Feir~\cite{BenF} predicted that periodic traveling-wave solutions to the Euler equations on an infinite-depth fluid are linearly unstable with respect to perturbations with long wavelengths.  Whitham~\cite{WhithamInstab} and Benjamin~\cite{B} predicted that periodic traveling-wave solutions to the Euler equations on a finite-depth fluid are unstable provided the nondimensional wavenumber, $k$, and nondimensional still fluid depth, $h$, satisfy $kh>1.363$.  Bridges \& Mielke~\cite{BridgesMielke} rigorously proved this finite-depth result.  These long-wavelength instabilities, in infinite- or finite-depth, are known as Benjamin-Feir or modulational or subharmonic instabilities.  

Longuet-Higgins~\cite{longuet1978instabilities} established that sufficiently steep periodic solutions of the Euler equations on an infinite depth fluid are linearly unstable with respect to perturbations of the same wavelength as the solution.  Instabilities that have the same period as the unperturbed solution are known as superharmonic or co-periodic instabilities.  Tanaka~\cite{tanaka1983stability} established that the transition from superharmonic stability to superharmonic instability on an infinite-depth fluid occurs when the Hamiltonian is maximized.  Saffman~\cite{saffman1985superharmonic} proved that this stability transition occurs when the Hamiltonian is maximized as a function of wave height (or steepness).  Zufiria \& Saffman~\cite{zufiria1986superharmonic} extended Saffman's result to the finite-depth case.

Recently, Korotkevich {\emph{et al.}}~\cite{SHInstabStokesWaves} and Deconinck {\emph{et al.}}~\cite{SteepStokes} examined the stability spectra of near-extreme periodic traveling-wave solutions to the Euler equations on an infinite-depth fluid.  The main purpose herein is to show that the Whitham equation, which was proposed as a small-amplitude approximation of the Euler equations on a finite-depth fluid, shares a striking number of properties with the Euler equations in the large-amplitude limit. 

\subsection{The Whitham equation}

Whitham~\cite{Whitham,Whithambook} proposed the following nondimensional model of waves on a shallow (finite-depth) fluid with horizontal bathymetry
\begin{equation}
    u_t+\mathcal{K}*u_x+\frac{3}{2}uu_x=0,
    \label{Whitham}
\end{equation}
where $u=u(x,t)$ represents the free-surface displacement of the fluid, $x$ is the horizontal coordinate, $t$ is the temporal coordinate, $*$ represents convolution, and $\mathcal{K}$ is the Fourier multiplier defined by
\begin{equation}
    \hat{\mathcal{K}}(k)=\sqrt{\frac{\tanh(k)}{k}}.
\end{equation}
All variables are nondimensional.  We assume $2\pi$-periodic boundary conditions in the $x$-dimension and that the fluid has a nondimensional undisturbed depth of $h=1$.  Equation (\ref{Whitham}) reproduces the {\emph{uni}}directional phase speed of the Euler equations for all wavenumbers.  It is known as the Whitham equation for water waves.  Moldabayev {\emph{et al.}}~\cite{Moldabayev} identify a scaling regime in which the Whitham equation can be derived from the Hamiltonian theory of surface water waves.  The Whitham equation has been shown to accurately model the evolution of long waves of depression in laboratory experiments, see for example~\cite{trillo,WhithamComp}.  Particular generalizations of the Whitham equation that allow for nonhorizontal bathymetry have been shown to accurately model the evolution of waves over nonhorizontal bathymetry~\cite{Bathy}.

The Whitham equation has three known conserved quantities: the solution mean,
\begin{equation}
    \mathcal{M}=\frac{1}{2\pi}\int_{-\pi}^{\pi}u~dx,
    \label{Mean}
\end{equation}
the $\mathcal{L}_2$-norm,
\begin{equation}
    \mathcal{L}_2=\int_{-\pi}^{\pi}u^2~dx,
    \label{L2norm}
\end{equation}
and its Hamiltonian,
\begin{equation}
    \mathcal{H}=\frac{1}{2}\int_{-\pi}^{\pi}\left( u\mathcal{K}*u+\frac{1}{2}u^3 \right)dx.
    \label{Hamiltonian}
\end{equation}
Note that the Euler equations also conserve the solution mean and the $\mathcal{L}_2$-norm of the surface displacement, along with its Hamiltonian (though the Hamiltonian for the Euler equations has a different form than that of the Whitham equation).

Ehrnstr\"om and Kalisch~\cite{EK} proved that the Whitham equation admits periodic traveling-wave solutions and computed some of these solutions.  Ehrnstr\"om \& Wahl\'en~\cite{WhithamCusp} proved that the Whitham equation has a solution with maximal wave height and that this solution is cusped.

Hur \& Johnson~\cite{HurJohnson2015} proved that small-amplitude periodic traveling-wave solutions of the Whitham equation are stable with respect to the modulational instability if $k<1.146$ and are unstable with respect to the modulational instability if $k>1.146$.  Sanford {\emph{et al.}}~\cite{Sanford2014} numerically corroborated this result and established that all moderate- to large-amplitude solutions are unstable with respect to the modulational instability.  Carter {\emph{et al.}}~\cite{WhithamSI} numerically studied the superharmonic instability in the Whitham equation and made comparisons with results from the finite-depth Euler equations.  They found that Whitham solutions steeper than $s=0.062$ are unstable with respect to the modulational instability and that solutions steeper than $s=0.104$ are unstable with respect to the superharmonic instability.  

The remainder of the paper is outlined as follows.  Section \ref{SectionTW} presents some $2\pi$-periodic traveling-wave solutions of the Whitham equation and a few of their properties.  Section \ref{SectionStab} contains a stability analysis of solutions to the Whitham equation, with a focus on near-extreme solutions.  Section \ref{SectionComparison} contains comparisons of the Whitham stability results with the Euler stability results.  Finally, Section \ref{SectionSummary} contains a summary of the main results presented herein.

\section{Traveling-wave solutions}
\label{SectionTW}

\begin{figure}
    \begin{center}
        \includegraphics[width=12cm]{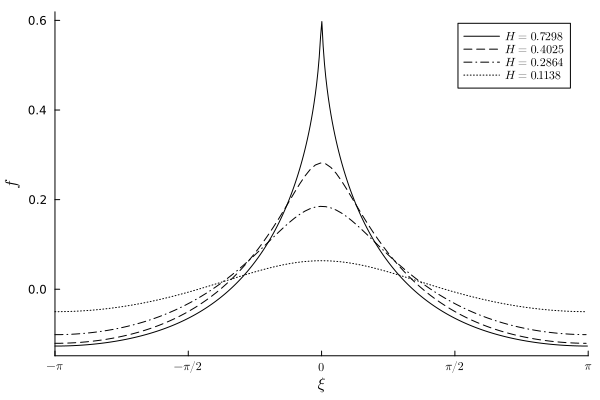}
        \caption{Plots of four zero-mean $2\pi$-periodic traveling-wave solutions to the Whitham equation.  The heights of the solutions are included in the legend.}
        \label{SolnsPlot}
    \end{center}
\end{figure}

\begin{figure}
    \begin{center}
        \includegraphics[width=12cm]{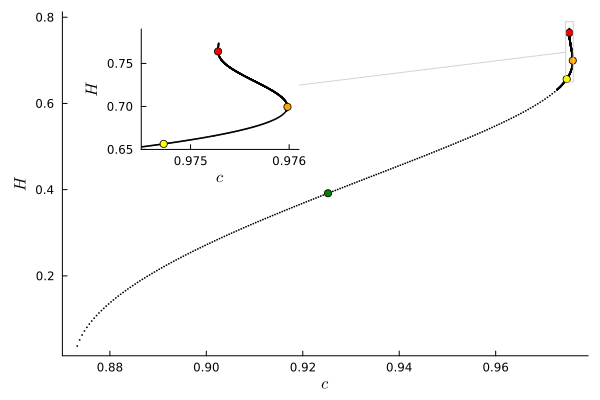}
        \caption{Plot of wave height versus wave speed for the branch of zero-mean $2\pi$-periodic solutions to the Whitham equation.}
        \label{BifPlot}
    \end{center}
\end{figure}

We compute $2\pi$-periodic traveling-wave solutions of the Whitham equation of the form 
\begin{equation}
    u(x,t)=f(x-ct)=f(\xi),
    \label{TWForm}
\end{equation}
using a Fourier basis and the branch-following method described in~\cite{ehrnstrom2013global,CVWhitham}.  Without loss of generality, we consider solutions with zero mean (i.e.~$\mathcal{M}=0$).  Figure \ref{SolnsPlot} contains plots of four zero-mean $2\pi$-periodic traveling-wave solutions of the Whitham equation.  Figure \ref{BifPlot} contains a plot of wave height, $H$, versus wave speed, $c$, for the branch of zero-mean $2\pi$-periodic solutions of the Whitham equation.  The values of the solution parameters at the colored dots are included in Table \ref{DataTable}.  Note that the curve has turning points at the orange and red dots.  The existence of the first of these turning points was first explored by Kalisch {\emph{et al.}}~\cite{WTurningPt}.  We hypothesize that the speed of the solutions continues to oscillate as the wave height increases.  However, we were unable to further extend the branch due to the large number of Fourier modes required to accurately resolve the nearly cusped solutions.  

\begin{figure}
    \begin{center}
        \includegraphics[width=12cm]{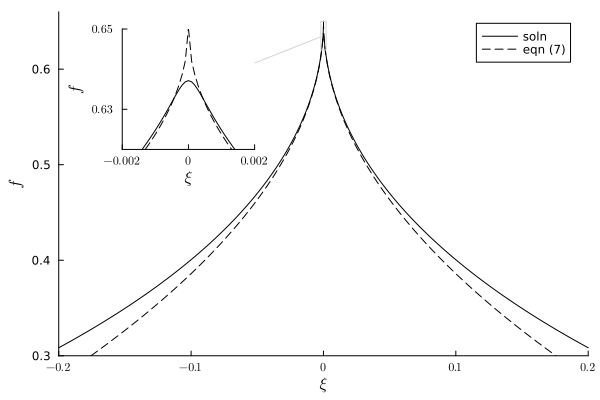}
        \caption{Plot comparing the crest of the solution corresponding to the red dot in Figure \ref{BifPlot} (solid curve) with the asymptotic form of the steepest solution near the peak given in equation (\ref{SteepestForm}) (dashed curve).}
        \label{NearlySteepestPlot}
    \end{center}
\end{figure}

Figure \ref{NearlySteepestPlot} contains a plot comparing the solution corresponding to the red dot in Figure \ref{BifPlot} with the asymptotic formula for the shape of the steepest solution derived by Ehrnstr\"om \& Wahl\'en~\cite{WhithamCusp} that is valid near the crest
\begin{equation}
    f\sim\frac{4}{3}\left(\frac{c}{2}-\sqrt{\frac{\pi}{8}}~|\xi|^{1/2}\right),\hspace{0.4cm}\text{as }|\xi|\rightarrow0.
    \label{SteepestForm}
\end{equation}
The two curves show good agreement in the vicinity of $\xi=0$.  The computed solution was obtained using $2^{16}$ Fourier modes and its parameter values are listed in Table \ref{DataTable}.  The inset plot demonstrates that the computed solution is smooth, while the solution with maximal steepness is not.

\begin{table}
    \begin{center}
      \begin{tabular}{c|ccccc}
        Color & $c$ & $H$ & $s$ & $\mathcal{H}$ & $\mathcal{L}_2$ \\
        \hline
        Green & 0.9252 & 0.3916 & 0.0623 & 0.0422 & 0.0940 \\
        Yellow & 0.9747 & 0.6564 & 0.1045 & 0.0691 & 0.1508 \\
        Orange & 0.9760 & 0.6995 & 0.1113 & 0.0685 & 0.1497 \\
        Red & 0.9753 & 0.7639 & 0.1216 & 0.0678 & 0.1482 \\
      \end{tabular}
      \caption{Parameter values for the solutions that correspond to the colored dots that appear in many of the figures.  The parameters $c$, $H$, $s$, $\mathcal{H}$, and $\mathcal{L}_2$ represent the wave speed, wave height, wave steepness, Hamiltonian, and $\mathcal{L}_2$-norm, respectively.}
      \label{DataTable}
    \end{center}
\end{table}

\begin{figure}
    \begin{center}
        \includegraphics[width=12cm]{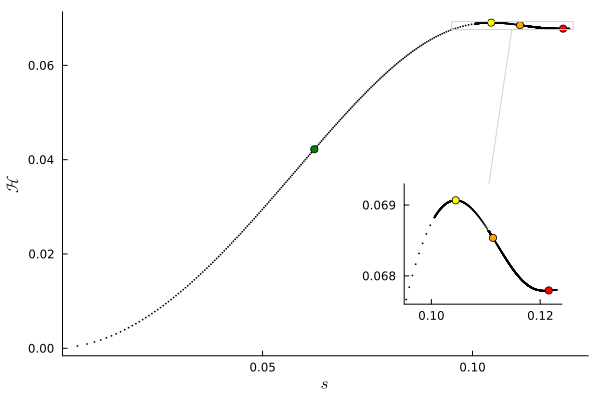}
        \caption{Plot of the Hamiltonian versus wave steepness for the branch of zero-mean $2\pi$-periodic traveling-wave solutions to the Whitham equation.}
        \label{HamVssPlot}
    \end{center}
\end{figure}

Figure \ref{HamVssPlot} contains a plot of the Hamiltonian, $\mathcal{H}$, versus the wave steepness for the branch of zero-mean $2\pi$-periodic traveling-wave solutions to the Whitham equation.  The Hamiltonian has local minima at the origin and at the red dot.  It has a local maximum at the yellow dot.  The yellow dot is the global maximum on the interval examined.  Under-resolved numerical simulations suggest that the Hamiltonian has a second local maximum after the red dot.  However, due to the resolution required near this wave height, we are not able to ensure the accuracy of this statement.  We hypothesize that the Hamiltonian continues to oscillate as wave steepness further increases.

\begin{figure}
    \begin{center}
        \includegraphics[width=12cm]{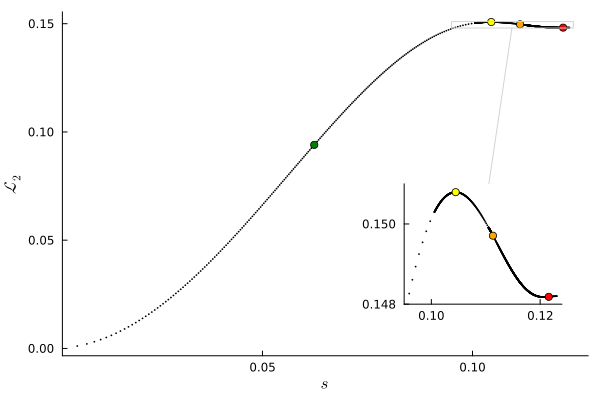}
        \caption{Plot of the $\mathcal{L}_2$-norm versus wave steepness for the branch of zero-mean $2\pi$-periodic traveling-wave solutions to the Whitham equation.}
        \label{L2VssPlot}
    \end{center}
\end{figure}

Figure \ref{L2VssPlot} shows that the plot of the $\mathcal{L}_2$-norm versus wave steepness is qualitatively similar to the plot of Hamiltonian versus wave steepness.  Just like the Hamiltonian, the $\mathcal{L}_2$-norm has local minima at the origin and at the red dot, and a local maximum at the yellow dot.  Under-resolved numerical simulations suggest that the $\mathcal{L}_2$-norm has a second local maximum after the red dot.

\begin{figure}
    \begin{center}
        \includegraphics[width=12cm]{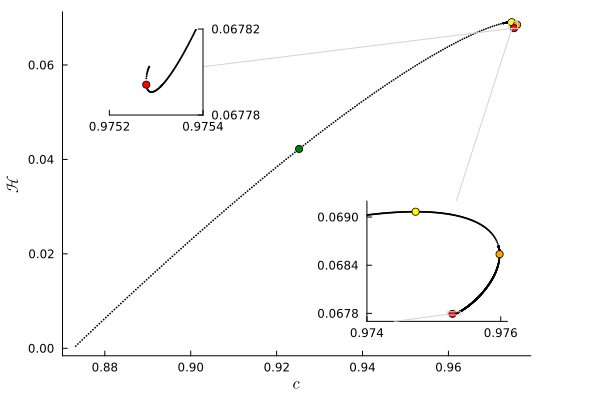}
        \caption{Plot of the Hamiltonian versus wave speed for the branch of zero-mean $2\pi$-periodic traveling-wave solutions to the Whitham equation.}
        \label{HamVscsPlot}
    \end{center}
\end{figure}

Figure \ref{HamVscsPlot} contains a plot of the Hamiltonian versus wave speed for the branch of zero-mean $2\pi$-periodic traveling-wave solutions of the Whitham equation.  This plot shows that the Hamiltonian has a global maximum at the yellow dot.  The curve has turning points at the orange and red dots.  The plot of the $\mathcal{L}_2$-norm versus $c$ is qualitatively similar to the plot of $\mathcal{H}$ versus $c$ and is therefore omitted.  These plots show that the solution with maximal speed (the orange dot) has steepness higher than the solution with maximal Hamiltonian and $\mathcal{L}_2$-norm (yellow dot).  Finally, note that the maximum of the Hamiltonian (yellow dot) occurs at a speed less than the maximal speed (orange dot).

\section{Stability}
\label{SectionStab}

In order to study stability of traveling-wave solutions to the Whitham equation, enter a coordinate frame moving with the speed of the solution by letting $\xi=x-ct$.  This converts the Whitham equation to
\begin{equation}
    u_t-cu_\xi+\mathcal{K}*u_\xi+\frac{3}{2}uu_\xi=0.
    \label{WhithamMoving}
\end{equation}
We consider perturbed solutions of the form
\begin{equation}
    u_{pert}(\xi,t)=u(\xi)+\epsilon v(\xi,t)+\mathcal{O}(\epsilon^2),
    \label{PertForm}
\end{equation}
where $u$ is a traveling-wave solution, $\epsilon$ is a small constant, and $v$ is the leading-order term of the perturbation.  Substituting (\ref{PertForm}) into (\ref{WhithamMoving}) and linearizing gives
\begin{equation}
    v_{t}-cv_{\xi}+\mathcal{K}*v_{\xi}+\frac{3}{2}\left(uv\right)_\xi=0.
    \label{LinearPDE}
\end{equation} 
We use the Fourier-Floquet-Hill method~\cite{DK} to solve this equation.  Without loss of generality, assume
\begin{equation}
    v(\xi,t)=\mbox{e}^{i\mu \xi}V(\xi)\mbox{e}^{\lambda t}+c.c.,
    \label{vForm}
\end{equation}
where $\mu\in [-\frac{1}{2},\frac{1}{2})$ is known as the Floquet parameter, $\lambda$ is a complex constant, $c.c.$~stands for complex conjugate, and $V(\xi)$ is a $2\pi$-periodic complex-valued function with Fourier series
\begin{equation}
    V(\xi)=\sum_{j=-N}^{N}\hat{V}(j)\mbox{e}^{i j\xi},
    \label{VForm}
\end{equation}
where $N$ is a large positive integer.  If $\mu=0$, then the perturbation has the same $\xi$-period as the unperturbed solution.  If $\mu\ne0$, then the perturbation has a period larger than that of the unperturbed solution.

Substituting (\ref{vForm}) and (\ref{VForm}) into (\ref{LinearPDE}) gives the following matrix eigenvalue problem that depends on the parameter $\mu$
\begin{equation}
    \hat{\mathcal{L}}\hat{\mathcal{V}}=\lambda\hat{\mathcal{V}},
    \label{EvalProb}
\end{equation}
where $\hat{\mathcal{V}}=(\hat{V}(-N),\hat{V}(-N+1),\dots,\hat{V}(N))^T$ and the elements of the matrix $\hat{\mathcal{L}}$ are given by
\begin{equation}
    \hat{\mathcal{L}}_{mn}=  \begin{cases} 
        i\left(\mu+m\right)\left(c-\hat{\mathcal{K}}\left(\mu+m\right)\right), & m=n, \\
        -\frac{3}{2}i\left(\mu+m\right)\hat{u}(n-m), & m\ne n,
     \end{cases}
    \label{LHat}
\end{equation}
where the $\hat{u}$'s are the Fourier coefficients of $u$.  Given a solution of the Whitham equation, we choose an equally-spaced sampling of $\mu$ values and find the corresponding eigenvalues and eigenvectors of $\mathcal{L}$ using a standard eigensolver (``eigen'' in the Julia programming language).  If there exists a perturbation with $\mu=0$ and $\Re(\lambda)>0$, then the solution is said to be linearly unstable with respect to the superharmonic instability.  If there exists a perturbation with $0<\mu\ll1$ and $\Re(\lambda)>0$, then the solution is said to be linearly unstable with respect to the modulational instability.  The solution is said to be spectrally stable if there exist no $\mu$ such that $\Re(\lambda)>0$.

\begin{figure}
    \begin{center}
        \includegraphics[width=12cm]{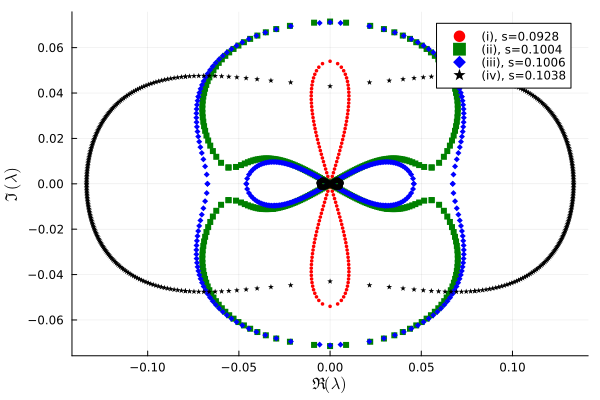}
        \caption{Stability spectra for four solutions to the Whitham equation with differing wave steepnesses before the onset of the superharmonic instability.}
        \label{StabPlot}
    \end{center}
\end{figure}

Solutions to the Whitham equation with steepness $s<0.062$ (and period $2\pi$) are stable.  This is consistent with the Hur \& Johnson~\cite{HurJohnson2015} result that small-amplitude solutions of the Whitham equation are stable with respect to the modulational instability when $k<1.146$.  When the steepness of the solutions surpasses $s=0.062$, they become unstable with respect to the modulational instability and their stability spectra contain a figure eight centered at the origin, see plot (i) in Figure \ref{StabPlot}.  This transition corresponds to the green dot in Figures \ref{BifPlot} and \ref{HamVssPlot}-\ref{HamVscsPlot}.  As wave steepness increases, the figure eight increases in size and morphs into ``mushrooms'' with lobes that bend and eventually intersect the $\Re(\lambda)$-axis away from the origin, see plot (ii) in Figure \ref{StabPlot}.  After the lobes intersect the $\Re(\lambda)$-axis, the spectra include a figure infinity centered at the origin surrounded by a vertical peanut-like shape, see plot (iii).  As the steepness increases further, the figure infinity decreases in size and the vertical peanut transforms into a horizontal peanut, see plot (iv).  These four solutions are all unstable, but are stable with respect to the superharmonic instability.

\begin{figure}
    \begin{center}
        \includegraphics[width=12cm]{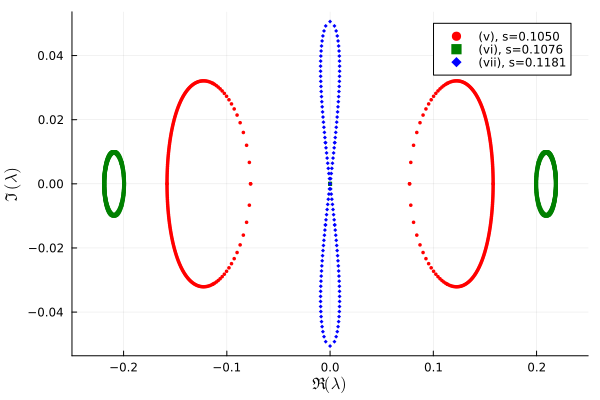}
        \caption{Stability spectra for three solutions to the Whitham equation with differing wave steepnesses above the onset of the superharmonic instability.  Plot (vii) also includes two small ovals centered near $\lambda=\pm 1.0976$ that are omitted so the finer details near the origin are visible.}
        \label{StabPlot2}
    \end{center}
\end{figure}

Solutions to the Whitham equation become unstable with respect to the superharmonic instability once their steepness surpasses $s=0.1045$.  In other words, solutions to the Whitham equation become unstable with respect to the superharmonic instability once their steepness is large enough that the first local maximum of the Hamiltonian (the yellow dot in Figure \ref{HamVssPlot}), or $\mathcal{L}_2$-norm (the yellow dot in Figure \ref{L2VssPlot}), is reached.  As this maximum is attained, the figure infinity collapses to the origin and the horizontal peanut pinches off into two oval-like shapes centered on the $\Re(\lambda)$-axis, see plot (v) in Figure \ref{StabPlot2}.  There is no figure eight or figure infinity in this spectrum.  The spectrum is confined to the imaginary axis other than the two ovals centered on the $\Re(\lambda)$-axis.  Regardless, all values of $\mu$ give instabilities as part of the ovals.  As the wave steepness continues to increase, the ovals decrease in size and move away from the origin while remaining centered on the $\Re(\lambda)$-axis, see plot (vi).  Eventually a new figure eight centered at the origin develops (in between the two ovals), see plot (vii).  Note that the ovals centered at $\lambda\approx\pm1.0976$ are omitted from plot (vii) so that the details near the origin are more apparent.  As the solution steepness further increases, the new figure eight undergoes transitions similar to those experienced by the first figure eight (i.e.~the figure eight that was created when the steepness reached $s=0.062$) while remaining between the two ovals centered on the $\Re(\lambda)$-axis that continue to decrease in size and move away from the origin.  When the local minima of the Hamiltonian and $\mathcal{L}_2$-norm at the red dots in Figures \ref{HamVssPlot} and \ref{L2VssPlot} are reached, a second superharmonic instability is created and the spectra are composed of four ovals centered on the $\Re(\lambda)$-axis.  Additional under-resolved simulations suggest that this pattern (the creation of a figure eight centered at the origin which bifurcates into two ovals once the solution becomes steep enough) is repeated as the steepness is increased beyond the local minimum of the Hamiltonian at the red dot.

\section{Comparison with results from the Euler equations}
\label{SectionComparison}

Although the Whitham equation was proposed as a simplification of the finite-depth Euler equations in the small-amplitude limit, it shares a surprising number of similarities to the Euler equations in the large-amplitude limit.  These similarities (and some differences) are discussed in this section.  The following comments assume that the solutions to both the Whitham and Euler equations have nondimensional period $2\pi$ and correspond to a nondimensional depth of $h=1$, except when noted.

\begin{enumerate}
    \item{For large values of wave steepness, the plot of the Whitham Hamiltonian oscillates as a function of wave steepness similarly to how the plot of the Euler Hamiltonian oscillates as a function of steepness.}
    \item{The Whitham solution with maximal speed has a steepness that is larger than the steepness of the solution corresponding to the maximum of the Hamiltonian.  This is similar to the Euler result.}
    \item{Small-amplitude solutions of the Whitham equation are stable while small-amplitude solutions to the finite-depth Euler equations are unstable.  Deconinck \& Oliveras~\cite{DeconinckOliveras} showed that small-amplitude solutions to the Euler equations are unstable with respect to ``high-frequency'' instabilities, but are stable with respect to the modulational instability.  The Whitham equation does not admit these high frequency instabilities.  Deconinck \& Trichtchenko~\cite{BernardOlga} showed that this apparent discrepancy is due to the unidirectional nature of the Whitham equation.}
    \item{Moderately steep solutions to both the Euler and Whitham equations are unstable with respect to the modulational instability.  The modulational instability onset for the Whitham equation occurs at $s=0.062$ while the onset in the Euler equations with dimensionless depth $h=1$ occurs at $s=0.085$, see~\cite{WhithamSI}.}
    \item{Solutions to both the Euler and Whitham equations with sufficiently large steepness are unstable with respect to superharmonic instabilities.  The superharmonic instability onset for the Whitham equation occurs at $s=0.1045$ while the onset in the Euler equations with dimensionless depth $h=1$ occurs at $s=0.099$, see~\cite{WhithamSI}.}
    \item{The onset of the superharmonic instability in the Whitham equation occurs at the first maximum of the Hamiltonian, just as in the finite-depth Euler equations, see~\cite{zufiria1986superharmonic}.}
    \item{The Whitham equation admits additional superharmonic instabilities each time an extremum of the Hamiltonian is achieved.  Similar behavior is seen in the infinite-depth Euler equations, see~\cite{SteepStokes}.  To the best of our knowledge, there are not yet any published related results for the finite-depth Euler equations.}
    \item{When the superharmonic instability is created, the most unstable mode corresponds to $\mu=\pm0.5$.  This means that the most unstable mode has a period that is twice that of the unperturbed solution.  It appears that the $\mu=\pm0.5$ mode remains the most unstable mode as the solution steepness increases.  This is different than the infinite-depth Euler result where the $\mu=0$ and $\mu=\pm0.5$ perturbations alternate between being the most unstable.}
\end{enumerate}

\section{Summary}
\label{SectionSummary}

We have shown that the Whitham equation, an equation proposed as a model for small-amplitude waves on shallow water, has a number of similarities with the Euler equations in the large-amplitude limit.  In particular, (i) the Hamiltonian oscillates as a function of solution steepness once the steepness is large enough and (ii) superharmonic instabilities are created at each nonzero critical point of the Hamiltonian.


\section*{Acknowledgements}
The author thanks Anastassiya Semenova and Bernard Deconinck for especially helpful conversations.  The author also thanks Malek Abid, Eleanor Byrnes, Sergey Dyachenko, Marc Francius, Henrik Kalisch, and Christian Kharif for additional helpful discussions.

\end{document}